\documentstyle[prl,aps,multicol,epsf]{revtex}

\begin{document}
\draft
\widetext

\title{Anderson localization due to spin disorder:\\
a driving force of temperature-dependent metal-semiconductor transition\\ 
in colossal-magnetoresistance materials}

\author{Eugene Kogan$^{}$\cite{emk}}
\address{Jack and Pearl Resnick Institute 
of Advanced Technology,\\
Department of Physics, Bar-Ilan University, Ramat-Gan 52900, 
Israel}

\author{Mark Auslender$^{}$\cite{em}}
\address{Department of Electrical and Computer Engineering\\
Ben-Gurion University of the Negev\\
P.O.B. 653, Beer-Sheva, 84105 Israel}

\date{\today}
\maketitle
\widetext
\begin{abstract}
\leftskip 54.8pt
\rightskip 54.8pt

We study temperature induced metal-insulator transition in doped
ferromagnetic semiconductors, described by s-d exchange model.
The transition is a result of the mobility edge movement, the
disorder being due to magnetic ions spin density fluctuations.
The electrons are described in the tight binding approximation.
Using ideas and methods of Anderson localization theory we obtain simple
formulas, which connect the mobility edge with
short-range order characteristics of the magnetic subsystem --
static spin correlators. We discuss the application of the theory to
several groups of  novel colossal-magnetoresistance materials and
include the reproduction of the paper \cite{kog1} published by us 10 years ago.  

\end{abstract}

\pacs{ PACS numbers: 71.28.+d, 71.30.+h, 72.15.Rn, 72.20.-i }

\begin{multicols}{2}

\narrowtext

\section{Pre-introduction}

Conducting ferromagnets exhibiting resistivity peak and  colossal magnetoresistance
 near the Curie point
such as europium oxide- and sulfide-based compounds EuX${}_{1-y}$, 
Eu${}_{1-x}$R${}_{x}$X where X=O, S, R=La, Gd and chromium-chalcogenide 
spinels ACr${}_2$X${}_4$ where A=Cd, Hg, Cu, Fe, X=S, Se were studied from 60's 
to late 70's extensively. Because of difficulties in sample preparation and
other reasons the interest of RD community in these 
compounds declined in early 80's. The same, even earlier, happened to manganite 
perovskites La${}_{1-x}$A${}_{x}$MnO${}_3$ where A=Ca, Ba, Sr, Pb. 
Nevertheless, some studies in this field continued through the 80's. In 1993, in 
connection with new thin-film technology for manganite perovskites 
\cite{vonhelm} the interest in these, called today colossal magnetoresistance 
(CMR), compounds has been revived. As a result,  approaches and concepts 
developed and applied
during 70's and 80's has again been brought 
into the focus of attention. Partial list includes: 
magnetic polaron versus FM-AFM phase separation,
ferromagnetism in narrow conduction band, non-pole structure of Green function 
in s-d and Hubbard models. This concerns also the concept of
localization driven by magnetic fluctuations

We  published 10 years ago a series of papers on the resistivity peak and 
magnetoresistive effect in magnetic semiconductors \cite{kog1,aus,kog2}. 
To the best of our 
knowledge, the present authors were the first to apply the concept of mobility 
edge for  disorder determined by spin fluctuation. The concept of
localization driven by 
magnetic fluctuations  has become actual again, when  the problem of 
resistivity peak and  colossal magnetoresistance has again been brought 
into a focus of attention.

Below we reproduce our paper "Anderson 
localization in ferromagnetic semiconductors due to spin disorder. 
I. Narrow conduction band" \cite{kog1}, the first one from this series.

\section{Introduction}

Magnetic semiconductors form a wide class of materials with
unique physical properties, strong temperature dependence of the
electroresistivity being one of them \cite{methfessel}. This phenomenon was
most thoroughly studied experimentally in cubic ferromagnetics,
such as CdCr${}_2$Se${}_4$, EuO and EuS. All these materials,
when being deficient of chalcogen (oxygen) have n-type
conductivity. The deficit being strong enough, at low
temperatures these materials have temperature-independent
metallic conductivity or conductivity with small activation
energy (a few meV); when the temperature $T$ approaches the Curie
temperature $T_c$ electroresistivity rises sharply. The rise can
amount to four orders of magnitude in CdCr${}_2$Se${}_4$,\cite{amith}
six orders in EuS${}_{1-x}$ \cite{shapira}, and even twelve orders in 
EuO${}_{1-x}$ \cite{oliver,torrance,shapira73}. The same behavior of the 
resistivity can
be seen in Eu${}_{1-x}$La${}_x$S \cite{methfessel}, 
Eu${}_{1-x}$Gd${}_x$O \cite{shoenes} 
and Eu${}_{1-x}$Gd${}_x$O${}_{1-y}$ \cite{samohvalov}. In the paramagnetic
region there are two types of behavior: in CdCr${}_2$Se${}_4$
and EuO${}_{1-x}$ (x<.4\%)the resistivity has large activation
energy ($\sim .2$ eV and $\sim .3$ eV respectively) in the whole
temperature range. At the same time the resistivity of 
EuS${}_{1-x}$ and of EuO and EuS being doped by a three-valent
rare earth metal (and also of Eu${}_{1-x}$Gd${}_x$Se, which at
$x>.01$ is ferromagnetic \cite{methfessel}) returns at hight temperatures
to the initial quasi-metallic behavior.

The common feature of all these materials is a strong interaction
between electrons and spins of the magnetic d- and f-ions. It is
commonly accepted now that the electrical of these materials 
are determined by the interaction of
electrons with the magnetic subsystem. Pre-existing explanations
of the above-mentioned resistivity behavior used the following
two approaches:

1. magnetopolaronic approach (in various modifications), which
was used for the description of metal-insulator transition in 
EuO${}_{1-x}$;

2. concentrational approach, when it was supposed that the
resistivity rise is due to the movement of the conductivity band
edge with the temperature relative to the fixed impurity levels.

We describe the conductivity behavior in all above-mentioned
materials in a novel approach, which is based upon the ideas and
methods of localization theory. The differences in the
resistivity behavior between materials, as it will be shown
later, are due to the different relations of the band width to
the spin exchange splitting and also to the different
electron-impurity interactions. 

\section{Hamiltonian and Theoretical Formulation}

Hamiltonian of the electron is
\begin{equation}
\label{ham}
H_e=\sum_{l,l',\sigma}V(l-l')c^{\dagger}_{l\sigma}c_{l'\sigma}
-2I\sum_{l,\sigma,\sigma'} {\bf S}_l{\bf \cdot s}_{\sigma,\sigma'}
c^{\dagger}_{l\sigma}c_{l\sigma'}+V_{\rm e-i},
\end{equation}
where V(l-l') is the hopping integral, I is the s-d(f) exchange
integral, ${\bf S}_l$ is spin localized on $l$ site, 
$c^{\dagger}_{l\sigma}, \; c_{l\sigma}$ are electron creation and
annihilation operators, and $V_{\rm e-i}$ is the potential of
electron-impurity interaction. (We suppose that magnetic and
crystallographic lattices coincide and that there is one electron
orbit per site).

It is convenient to separate out two limiting cases \cite{nagaev}

1. narrow conduction band $W\ll 2|I|S$, where $W$ is the band width;

2. wide conduction band $W\gg 2|I|S$.

As it was shown in band calculations, the conduction band in CdCr${}_2$Se${}_4$, 
formed by p-d$\gamma$ orbitals (3d$\gamma$ Cr + 4p Se) is narrow \cite{oguchi}, 
and p-d$\epsilon$ conduction bands in EuX are wide \cite{cho}. It is worth 
noting however, that strong inequalities for these materials do not hold.

In Part I of our paper we shall be interested in the narrow band
case. In this case it is convenient to consider Hamiltonian (\ref{ham}) in the 
representation which diagonalizes the main term -- s-d exchange. We restrict 
ourselves to the quasiclassical limit $2S\gg 1$, when we can consider spins as 
vectors with fixed length,
\begin{equation}
{\bf S}_l=S{\bf n}_l;
\end{equation}

The transformation of the initial creation and annihilation operators 
$c_{l\sigma}^{\dagger}, c_{l\sigma}$ into new ones $C_{l\sigma}^{\dagger}, 
C_{l\sigma}$ is equivalent to the quantization of the electron spin in the local 
reference frame, the OZ axis being parallel to ${\bf n}$. We have
\begin{equation}
\label{trans}
\left(\begin{array}{c}C_{l+}\cr C_{l-}\end{array}\right) 
=\left(\begin{array}{cc}u_{l}&v_l\cr 
-v_{l}\exp(i\phi_l)&u_{l}\exp(-i\phi_l)\end{array}\right) 
\left(\begin{array}{c}c_{l+}\cr c_{l-}\end{array}\right),
\end{equation}
where $u_l=\cos(\theta_l/2), v_l=\sin(\theta_l/2)\exp(-i\phi_l)$. Hamiltonian 
(\ref{ham}) after the transformation (\ref{trans}) takes the form
\begin{eqnarray}
\label{hamnew}
H_e=\sum_{\begin{array}{c} <ll'>\cr \alpha \alpha' \end{array}}
V^{\alpha \alpha'}(l,l')C^{\dagger}_{l\alpha}C_{l'\alpha'}+ \nonumber\\
\sum_{l\alpha \alpha'} S I^{\alpha \alpha'}C^{\dagger}_{l\alpha}C_{l\alpha'}
+V_{e-i},
\end{eqnarray}
where matrices (in spin space) $\hat{V}(l,l')$ and $\hat{I}$ are
\begin{eqnarray}
\hat{V}(l,l')&=&V\left(\begin{array}{cc}u_lu_{l'}+v_lv_{l'}^{\ast} &
 -u_lv_{l'} + v_lu_{l'}\cr
u_lv_{l'}^{\ast}-v_l^{\ast}u_{l'} &u_lu_{l'}+v_l^{\ast}v_{l'}\end{array}\right)
\nonumber \\
\hat{I}&=&\left(\begin{array}{cc}I&0\cr 0&-I\end{array}\right),
\end{eqnarray}
and nearest neighbors approximation is used.

The conduction band being narrow, we must retain in Eq. (\ref{hamnew}) the 
creation and annihilation operators with $\alpha=sgn I=+1$ only. The Hamiltonian 
then takes the form
\begin{eqnarray}
\label{hamef}
H_e= \sum_{<ll'>}V_{ll'} C_{l+}^{\dagger}C_{l'+}+V_{e-i}, \nonumber\\
V_{l,l'}=V({\bf n}_l,{\bf n}_{l'})=V\left[\cos\left(\frac{\theta_l}{2}\right) 
\cos\left(\frac{\theta_{l'}}{2}\right)+ \right .\nonumber\\
\left .\sin\left(\frac{\theta_l}{2}\right) 
\sin\left(\frac{\theta_{l'}}{2}\right)\exp i(\phi_l-\phi_{l'})\right]
\end{eqnarray}

When there is a complete order $(T=0)$ $\theta_l=0$. When $T>0$ the angles 
$\theta$ and $\phi$ and hence the Hamiltonian $H_e$ are the functions of time. 
Luckily for us there is a small parameter ratio of typical fluctuations 
frequency to electron energy $\sim kT_c/W$. Then the adiabatic approximation is 
applicable, and what we need to solve is static problem with the random 
Hamiltonian $H_e(t)$. Hence we obtain the disordered system with an off-diagonal 
disorder \cite{ziman}. We suppose also that the influence of the 
electron-impurity interaction on the conduction electrons is much less than the 
influence of magnetic disorder, and completely neglect the former.

The analysis of Hamiltonian (\ref{hamef}) is conveniently started with the 
density of states.    
\begin{figure}
\epsfxsize=2.5truein
\centerline{\epsffile{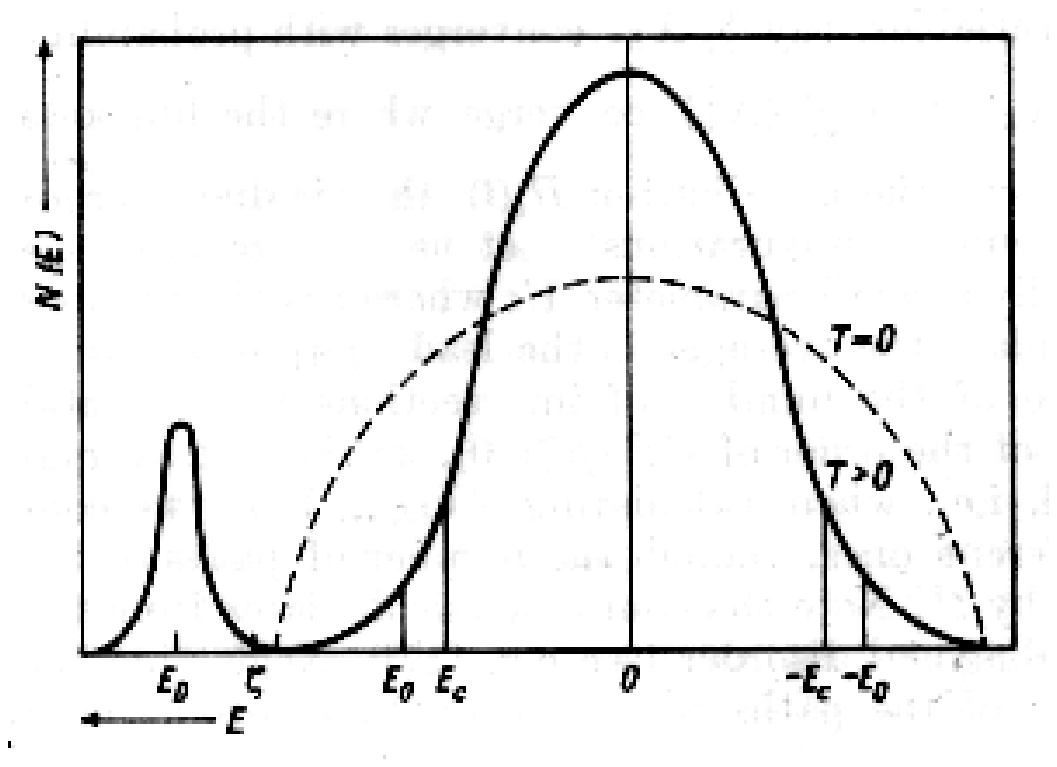}}
\label{Fig.1}
\end{figure}
%\vskip -11cm
\noindent
{\footnotesize {\bf FIG. 1.} 
The temperature dependence of the density of states in doped ferromagnetic 
semiconductor (the lower local band). $E_D$ is the impurity level, $\zeta$ is 
the chemical potential, $E_0$ is the band edge in the effective medium 
approximation, $\pm |E_c|$ are the mobility edges} 
\vskip .5cm

When some approximation of the effective medium is used, the influence of the 
disorder consists in the temperature dependent narrowing of the band. Such a 
conception was used to explain the temperature dependence of the 
electroresistivity of CdCr${}_2$Se${}_4$ \cite{amith}. It was supposed that the 
electric current is carried by the electrons activated from fixed impurity 
levels, 
created by $V_{e-i}$ to the red-shifted band edge. For the electrical 
resistivity the following formula was used
\begin{equation}
\label{rhodelta}
\rho(T)=\rho_0\exp\left[\frac{\Delta}{kT}\right]
\end{equation}
 where the activation energy $\Delta$ is defined by
\begin{equation}
\label{dif}
\Delta=E_0-E_D,
\end{equation} 
where $E_0$ is the band edge and $E_D$ is the impurity level. There exists a 
number of facts, however, which do not agree with this picture. In 
CdCr${}_2$Se${}_4$ above the temperature $T\simeq 200$ K a large blue shift of 
the band edge is observed \cite{methfessel}, but there is no activation energy 
change in this temperature range. Also a similar resistivity transition takes 
place in n-CdCr${}_2$S${}_4$ \cite{nikiforov}, where the band-edge shift is 
blue at all temperatures \cite{methfessel}. What is more important, according to 
modern conceptions for conductivity in disordered systems, it is the mobility 
edge $E_c$ which is important. We remind that the mobility edge is the energy 
which separates the localized from the extended states. For the calculation of 
the resistivity it is possible to use Eq. (\ref{rhodelta}), \cite{mott} but Eq. 
(\ref{dif}) should take the form
\begin{equation}
\Delta=E_c-\zeta,
\end{equation} 
where $\zeta$ is the chemical potential. As in Ref. \cite{amith} we shall 
consider $\zeta$ as temperature independent. Than our main task is to calculate 
$E_c$.
  
\section{The Mobility Edge Calculation}

Let us turn to Hamiltonian (\ref{hamef}). As is well known, the
problem of the mobility edge calculation is equivalent to the
calculation  of the convergence radius for a non-averaged Green
function self-energy perturbation series expansion. The site
diagonal representation of matrix elements of the Green function
being written in the form
\begin{equation}
\label{green}
G_{ll}(E)=\left[E-M_l(e)\right]^{-1},
\end{equation}
we obtain for the self-energy $M_l(E)$ the series
\begin{eqnarray}
M_l(E) = \frac{1}{E}\sum_{l_2} V_{ll_2}V_{l_2l}+\frac{1}{E^2}\sum_{l_2l_3} 
V_{ll_2}V_{l_2l_3}V_{l_3l}+\dots \nonumber\\
=\sum_{L}X_L.
\label{self}
\end{eqnarray}
Every term in $X_L$ corresponds to a definite path with $L$ steps
starting from the site $l_1$ and ending at it. To each step from
a site $l$ to one of its nearest neighbors $l'$ there
corresponds the factor $V_{ll'}$ and the passage of every
intermediate site produces multiplication by the factor $1/E$.
\cite{lichiardello}

It is worth noting that the analysis of the series (\ref{self}) is
much easier than that of the series involved in the case of the
traditional Andersen model with diagonal disorder \cite{ziman}. It is
due to the fact that the expectation value of every term in
Eq.(\ref{self}) is finite and, hence, the convergence of this series
with probability one is equivalent to its convergence in some
norm. Indeed we can use the theorem according to which the series
of independent random variables, say $\sum_L X_L$, converges with
probability one if and only if the series $\sum_L <X_L>$ and 
$\sum_L <X_L^2 >$ converge, where the brackets mean the average
\cite{feller} (in our case over the distribution $H_e(t)$, that is due
to the ergodic hypothesis the average over all spin
configurations). Let us analyze now the averaged series. We
introduce formally a small parameter $1/z$, where $z$ is the
number of nearest neighbors, and then calculate the averages in
the leading approximation with respect $1/z$. Since the ratio of
the number of intersections to the total number of sites for a
typical path is of the order of $1/z^p$ ($p > 0$), we shall on
averaging neglect the intersections at all, {\it i.e.}, when
calculating $<V_{l_1l_2} \dots V_{l_Ll_1}>$ we consider all the sites involved 
as different ones, though the number of paths of of the given Length $L$ being 
approximated by $z^L$ Note that our analysis deviates from the traditional one 
for the case of the off-diagonal disorder  (see e.g. Ref. \cite{antoniou}). 
Neglecting the correlations due to self-intersections of the paths we, however, 
take into account the correlations created by magnetic short-range order, which 
is more important in the problem.

We are interested in the convergence radius of the series
\begin{equation}
\label{ser}
<M(E)>= E\sum_L \frac{z^L}{E^L}M_L,
\end{equation}
where the following notation is introduced:
\begin{eqnarray}
M_L= <<V_{l_1l_2}V_{l_2l_3}\dots V_{l_Ll_1}>>= \nonumber\\
<\int d{\bf n}_1\dots d{\bf n}_L V({\bf n}_1,{\bf n}_2)\dots  \nonumber\\ 
V({\bf n}_L,{\bf n}_1)P({\bf n}_1,\dots  {\bf n}_L)>_L
\end{eqnarray}
with $P({\bf n}_1,\dots {\bf n}_L)$ being the probability density of a given 
spin configuration and the brackets $<\dots>_L$ denote the average over all 
paths of a given length. It is worth noting that in our approximation the 
convergence of the series $\sum_L<X_L>$ is equivalent to the convergence of 
$\sum_L<X_L^2>$.

Let us consider two limiting cases:

1. $T=0$

In this case the effective hopping integral does not fluctuate and the series 
take the form 
\begin{equation}
<M(E)>= E\sum_L \frac{z^LV^L}{E^L},
\end{equation}
and its convergence radius $|E_c|=zV$ naturally coincides with the exact band 
edge (the center of the band corresponds to $E=0$).

2. $T=\infty$.

In this case all the paths of a given length are identical, all the spin 
configurations equally participate and averaging is easy to make up
\begin{eqnarray}
M_L= \frac{1}{(4\pi)^L}
\int d{\bf n}_1\dots d{\bf n}_L V({\bf n}_1,{\bf n}_2)\dots  
V({\bf n}_L,{\bf n}_1)
\end{eqnarray}
This can be rewritten in the form
\begin{eqnarray}
M_L= \mbox{Tr}\{\hat{V}^L\},
\end{eqnarray}
where $\hat{V}$ is an operator with the kernel $V({\bf n},{\bf n}')/4\pi$. Then 
the convergence of the series (\ref{ser}) is equivalent to the convergence of 
$\sum_{L}(z\lambda)^L/E^L$, where $\lambda$ is the largest eigenvalue of the 
operator $\hat{V}$. Thus for the mobility edge we obtain 
\begin{equation}
\label{modedge}
|E_c|=z\lambda
\end{equation}
The eigenvalues of $\hat{V}$ are easy to obtain due to its factorizability, and 
the result is 
\begin{equation}
\label{modedge2}
|E_c|=\frac{zV}{2}
\end{equation}

The later case prompts the convergence analysis method for
arbitrary temperature. Let us use for the spin configuration
probability the chain superposition approximation due to Kirkwood
\cite{balesku}
\begin{equation}
\label{kirk}
P({\bf n}_1,\dots {\bf n}_L)=
\frac{P({\bf n}_1,{\bf n}_2)\dots P({\bf n}_L,{\bf n}_1)}
{P({\bf n}_1)\dots P({\bf n}_L)} 
\end{equation}
where $P({\bf n}_1,{\bf n}')$ is the pair probability and $P({\bf n})$ is the 
one-site probability. In this approximation all the
paths are again identical and we again obtain Eq.(\ref{modedge}), only
now $\lambda$ is the largest eigenvalue of the operator $\hat{V}$
with the kernel
\begin{equation}
\label{kernel}
\hat{V}({\bf n},{\bf n}')=\frac{V({\bf n},{\bf n}')
P({\bf n},{\bf n}')}{\left[P({\bf n})P({\bf n}')\right]^{1/2}}.
\end{equation}
Due to the variational principle $\lambda$ is the maximal value
of the functional
\begin{equation}
\label{functional}
F\{f({\bf n})\}=\frac{\int\int f^{\ast}({\bf n})\hat{V}({\bf n},{\bf n}')
f({\bf n}')d{\bf n}d{\bf n}'}{{\frac{1}{4\pi}\int|f({\bf n})|^2d{\bf n}}}.
\end{equation}
For the estimation of $\lambda$ we may use the trial function in the same form 
as in the molecular field approximation. In this approximation 
$P({\bf n},{\bf n}')=P({\bf n})\cdot P({\bf n}')$ and the operator $\hat{V}$ is 
again factorizable. The eigenfunction of $\hat{V}$ corresponding to the larger 
of the two eigenvalues has the form
\begin{equation}
f({\bf n})=[P({\bf n})]^{1/2}\cos(\theta/2).
\end{equation}
Using this trial function in the variational principle we obtain
\begin{eqnarray}
\label{edge}
|E_c|=\frac{zV}{2}\cdot \frac{1+2<n^z>+<{\bf n}_0\cdot{\bf n}_1>}
{1+<n^z>}= \nonumber\\
\frac{zV}{2}\cdot \frac{1+2<S^z>/S+<{\bf S}_0\cdot{\bf S}_1>/S^2}
{1+<S^z>/S}.
\end{eqnarray}
Hence the result is the mobility edge expressed in terms of magnetization and 
pair correlation function of the nearest-neighbor spins. Our method of mobility 
edge evaluation can be generalized also to the case of arbitrary relation 
between $IS$ and $W$ \cite{kogan3}.  For arbitrary temperature the calculation 
of the mobility edge is equivalent to solving the cumbersome integral equation. 
For $T\gg T_c$, however, this equation can be easily solved; the mobility edge 
then is 
\begin{equation}
\label{edgeapprox}
|E_c|=\frac{W}{4}+IS-\left[(IS)^2+\frac{W^2}{16}\right]^{1/2}.
\end{equation}

\section{Comparison with Experiment}

Though Eq.(\ref{edge}) is obtained for classical spins we'll use
it to explain the temperature dependence of the resistivity of 
CdCr${}_2$Se${}_4$ (Fig. 2). The temperature dependence of the
resistivity at low temperatures ($T < 50$ K) shows that the
chemical potential lies in the close vicinity of the band edge.
At those temperatures the resistivity is defined by the exact
position of $\xi$ relative to the band edge; at higher
temperatures when $E_c$ moves upward, the dependence of $\rho$ on
$T$ is determined mainly by the shift of $E_c$ and for the
chemical potential we may use the approximation $\xi=-zV$.
Inserting into Eq. (\ref{rhodelta}) the expression for the mobility edge
we obtain
\begin{equation}
\label{rho}
\rho=\rho_0\exp\left[\frac{1-<{\bf S_0\cdot S_1}>/S^2}{1+<S^z>/S}
\left(\frac{W}{4kT}\right)\right]
\end{equation}
where $W=2zV$ is the conductivity band width. Using for the correlator 
$<{\bf S}_0\cdot{\bf S}_1>$ the results of Ref. \cite{callen}, and for $<S^z>$ 
the molecular field approximation $(S=3/2)$, and choosing the appropriate values 
of $W$ and $\rho_0$, we obtain the plot presented on Fig.3.  

\begin{figure}
\epsfxsize=1.7truein
\centerline{\epsffile{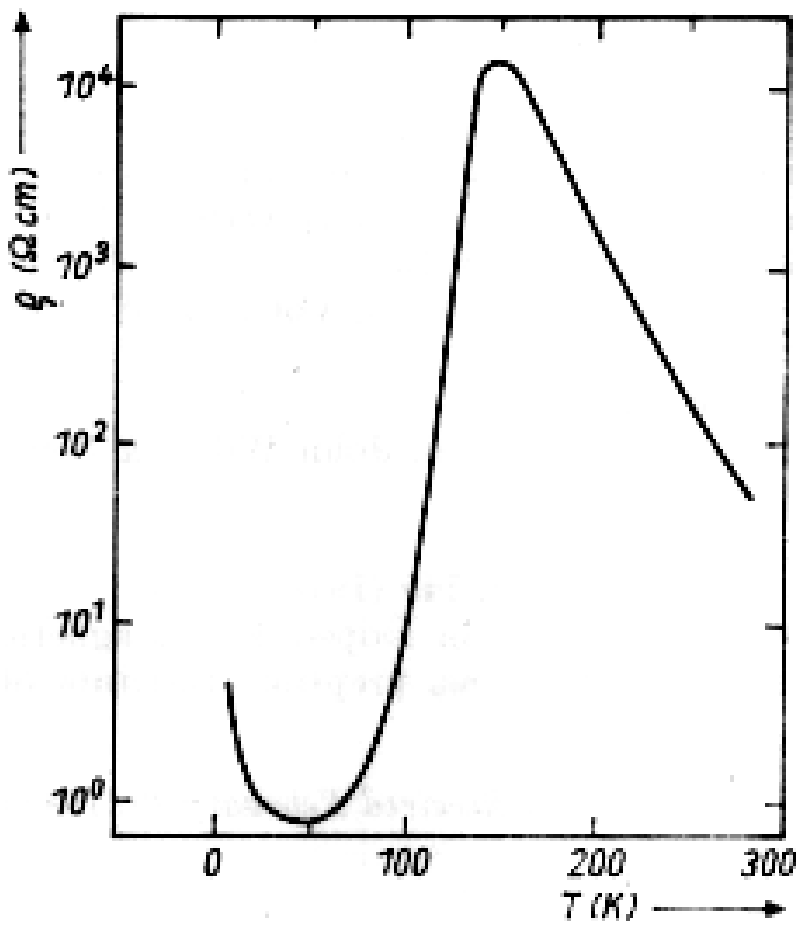} \epsfxsize=1.7truein \epsffile{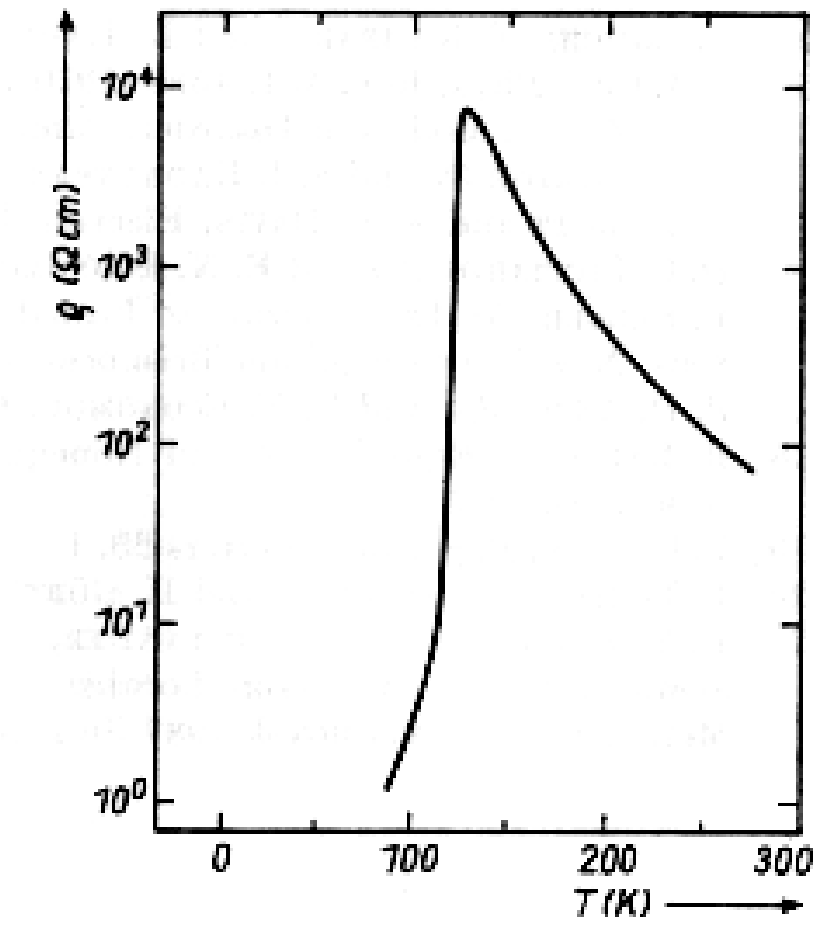}}
\end{figure}
\hskip 1.5cm Fig.2 \hskip 3.5cm Fig.3
\vskip .5cm

\noindent
{\footnotesize {\bf FIG. 2.} 
Electrical resistance of Cd${}_{0.99}$In${}_{0.01}$Cr${}_2$SE${}_4$ (reproduced 
from Ref. \cite{amith}.} 

\noindent
{\footnotesize {\bf FIG. 3.} 
Electrical resistance calculated on the basis of Eq. (\ref{rho}).} 
\vskip 1cm

The chosen value for the band width is $W=0.8$ eV, which well agrees with other 
estimations \cite{nagaev,oguchi}. We can deduce that Eq. (\ref{rho}) can claim 
for a quantitative description of the experimental dependence. This formula also 
explains the colossal negative magnetoresistance of the materials under 
consideration, as the magnetic field strongly influences the spin correlators in 
the exponent, especially near $T_c$.

In Refs. \cite{amith,treitinger} the data for high resistivity samples of 
CdCr${}_2$Se${}_4$ was presented. In this case the resistivity as the function 
of temperature has no maximum, but the dependence of $\log \rho$ on $1/T$ 
changes its slope after the transition to the paramagnetic region. Such a 
behavior is also described by Eqs. (\ref{rhodelta}) and (\ref{edge}) with the 
chemical potential $\zeta<-zV$.

\section{Discussion (10 years after)}

It is expedient to discuss the applicability of the approach  
of this ancestor paper to new classes of materials and its
place among more recent theoretical approaches. Our
principal point was that strongly interacting system of electrons and 
magnetic ions (s-d exchange model) in classical approximation 
for the ions spins and narrow electron band, reduces to a random-bond 
model with off-diagonal disorder depending on an instant spin 
configuration. This double-exchange (DE) Hamiltonian (Eq. (\ref{hamef}) in the 
present manuscript) has been reopened quite recently \cite{mhardag}.

The model, we believe, can be applied with minimal changes to
manganite pyrochlores 
discovered recently \cite{ramirez1,ramirez2}. Its  application to high-
carrier concentration CMR compounds such as manganite perovskites
demands discussions and probably additional ideas.
Formally, in this approximation the 
difference between, say, 
LaSrMnO$_{3}$ and CdCr$_{2}$Se$_{4}$ (except for carriers concentration) is 
that in the latter the doping electrons appear (due to Se vacancies) in the 
empty d-band, whereas in the former the doping holes appear in the filled 
d-band (due to capture of d-electrons by Sr atoms).
According to our concept, it is the movement of mobility edge, with 
temperature and magnetic field, relative to the Fermi level that explains 
the resistivity peak and CMR effect. It is concordant with the recent 
conclusion of Varma \cite{varma} who suggested the decisive role of the 
above spin-dependent disorder in metal-semiconductor transition and CMR 
effect in manganite perovskites. 

The viewpoint that spin disorder alone 
can not localize 
carriers at temperatures close to $T_{c}$ and 
in paramagnetic state has been argued by Millis et al. 
\cite{millis} on the basis of a perturbational calculation. 
Later this group presented very 
subtle arguments concerning probability distribution of spin-dependent hopping 
integral \cite{millis1} in favor of their conclusion. The matter is that, as 
with every random-bond model, DE Hamiltonian keeps a large portion of the states 
at the band center delocalized irrespective of how the bond fluctuations are 
strong. Therefore, it may turn out that the weight of the localized states be 
insufficient to embed large number of holes (0.2 to 0.3 per cell). After 
numerical simulation with the use of one-parameter scaling localization theory
\cite{li} it has been found that this is really the case for pure DE mechanism.
Note that for SC lattice our result of 1988 $E_{c}=-3V$ (see manuscript above)  
is in a reasonable agreement with the new more exact $E_{c}=-3.55V$ \cite{li}.

Thus, an agreement with experiment for manganites may be achieved by introducing 
some additional mechanism. Millis et al. \cite{millis1} developed and advocated 
Jahn-Teller polaron mechanism to explain the resistivity peak and CMR. On the 
other hand, in 1988  we proposed the mechanism based on the pinning of the Fermi  
energy by the donor levels, due to very weak overlap of the vacancy states with 
magnetic d-states in CdCr$_{2}$Se$_{4}$(see present paper). Recently our 
mechanism has been recognized \cite{ramirez1,ramirez2}, and applied to 
LaBaMnO$_{3}$ by Bebenin et al.\cite{beb}. In the case of manganite perovskites 
one may speculate that the Fermi level is more or less fixed by impurity band 
formed by Sr, Ba, etc. acceptor states. The band emerges in Mn d-band at low 
temperatures and holes are mobile because the d-band is delocalized. With 
increasing the temperature the mobility edge shifts towards higher energies and 
holes remain in localized part of the band. 

Recently an approximate pinning of the Fermi level has been found in a model 
where diagonal disorder is added to DE hopping Hamiltonian \cite{sheng}. 
Numerical simulation using one-parameter scaling localization theory have 
shown that the states at the Fermi energy are localized in PM region if the 
intensity of diagonal disorder is as intense as about the bandwidth 
\cite{sheng}. 
Though this model is very attractive, the physical origin of such strong 
fluctuation of Mn d-levels is not evident. Sheng et al. \cite{sheng} attribute 
it to potential fluctuations due to random substitution of La by Sr, Ca  etc. 
using experimental data on resistivity of polycrystalline samples. However, 
contrary to the assumption of \cite{sheng} there exist single crystals with 
zero-temperature resistivity smaller than  $10^{-4} \Omega \cdot cm$ which do 
exhibit large resistivity peak \cite{arkh}. This is an evidence in favor of weak 
substitutional disorder, and impurity band formation.

The mechanism of conduction in PM region is not established until now but  
it lies out of scope of this discussion. We hope to return to discussion on 
this problem in a forthcoming paper.

\end{multicols}

\end{document}